# Dynamics of local photoconductivity in GaAs and InP investigated by THz SNOM

Tinkara Troha, Arvind Singh, Petr Kužel, and Hynek Němec

Institute of Physics of the Czech Academy of Sciences, Na Slovance 2, 18200 Prague, Czech Republic

Terahertz scanning near-field optical microscope (THz-SNOM) is employed to measure ultrafast evolution of THz conductivity spectra after photoexcitation of GaAs and InP wafers using ultrashort laser pulses. Unlike in GaAs, the THz photoconductivity decay in InP is controlled mainly by the diffusion of electrons away from the photoexcited area, and also by the drift due to band-bending at the surface of the semiconductor. We propose and discuss several general strategies of the analysis of signals measured using THz-SNOM, and we estimate the accuracy of the obtained near-field photoconductivity spectra.

## 1. Introduction

Terahertz (THz) radiation is an efficient probe of charge transport mechanisms and properties in a large variety of materials, including classical semiconductors [1, 2], semiconductor nanostructures [3] and carbon-based materials [3, 4]. There are numerous advantages of using THz radiation: its strong interaction with free charges, the presence of characteristic spectral behavior due to carrier scattering rates falling into the THz spectral range, and non-contact nature of the interaction, which eliminates the possible electrode effects. Experiments using time-domain THz spectroscopy can naturally incorporate an optical branch for sample photoexcitation: using the (time-resolved) optical pump – THz probe experiments, it is then possible to follow the photoinduced conductivity with sub-picosecond time resolution [1, 5].

Nowadays, relevant experimental methods as well as data analysis are well established for the case of conventional (far-field) transmission spectroscopy; however, its principal limitation in the context of nanostructures is the long wavelength of THz radiation (1 THz ≙ 0.3 mm) which does not allow examination of individual deeply submillimeter objects due to the diffraction limit. This limit may be overcome using scanning near-field optical methods [6] or scanning tunnelling microscopy [7]. Particularly useful is a technique based on the detection of the radiation scattered from an oscillating tip (scattering-type SNOM): thanks to the strong enhancement of the incident field at the nano-sized tip, the scattered radiation encodes information about the local properties of the sample in the close vicinity of the tip apex [8, 9]. It is thus possible to examine the conductivity or photoconductivity with sub-micron spatial resolution even in the THz range [10, 11].

While the SNOM methods are frequently used for quantitative imaging without spectral resolution, the scattered broadband THz radiation contains in principle rich information about the local conductivity or permittivity spectra. The relation between the amplitude and phase of the scattered radiation and the sample properties is far from being simple [9] and the models developed so far inevitably involve considerable simplifications regarding the tip, the sample and their near-field interactions with the THz field. A point dipole model [12] is the simplest model which represents the tip as a finite size sphere placed close to a half-space sample. The scattered signal is than considered to be proportional to the polarizability of the sphere/sample system. For more quantitative analysis, the so-called finite dipole model was developed, both for a bulk sample [13] and for layered systems [14]. This model approximates the tip as a finite conducting spheroid. In real materials, deciphering of the sample properties can be considerably complicated by a possible spatial (time dependent photo-)conductivity profile along the surface normal caused, for example, by a band bending and the consequent depletion or accumulation of charges close to the semiconductor surface [15].

In this paper, we report on the measurements of the ultrafast time evolution of photoinduced THz conductivity spectra in GaAs and InP. These are two prototypical semiconductors with well-known properties and quite different photocarrier lifetimes, thus allowing us to verify the outcomes of the SNOM measurements. Despite the apparent simplicity of the investigated samples, we demonstrate that analysis of the THz SNOM spectra brings new insights compared to conventional transmission spectroscopy: namely, information on the depth profile can be revealed.

## 2. Experimental

The measurements are done with scattering type scanning near-field microscope (Neaspec) where a time-domain terahertz spectrometer operating in the spectral range between 0.5 – 2 THz is integrated into an atomic force microscope (AFM) device. The angle of incidence of *p*-polarized probing THz pulses is 60°. The AFM operates in the tapping mode; we use 40 nm platinum-iridium tips (Rocky Mountain Nanotechnology, 25PtIr200B-H40) driven at frequencies $\Omega_{tip}$ between 50 and 80 kHz. The tip-tapping amplitude was set to 150 nm and the minimum tip-sample distance was 1 nm. Femtosecond laser pulses (central wavelength 780 nm, maximum pulse energy 2 nJ, repetition rate 100 MHz, pulse duration 100 fs) synchronized in time with the probing THz pulses are used to photoexcite the sample. The scattered THz radiation is collected in a direction perpendicular to the incidence plane and at 60° with respect to the sample normal; the detected signal is demodulated at 2[nd] – 5[th] harmonics of the tip tapping frequency $\Omega_{tip}$. The scattered THz spectra were measured as a function of the pump-probe delay.

The investigated GaAs and InP are standard commercial semi-insulating wafers with optically polished surfaces. The measurements are performed under a dry air atmosphere.

# 3. Theory

## 3.1. Measurable signals

We will follow here the finite dipole model [13]. The scattered signal $s^{(m)}$ is assumed to be proportional to the effective tip polarizability $\alpha^{(m)}$ describing the near-field interaction between the tip and the sample, and to the far-field reflectance at the air-sample interface:

$$s^{(m)} \propto \alpha^{(m)} (1+r)^2, \tag{1}$$

where *m* denotes the harmonics of the tip tapping frequency (we concentrate namely on *m* = 2 and 3, in our analysis). The expression for the effective polarizability of the tip in the vicinity of a layered structure was derived in [14]. The second term on the right-hand side accounts for the fact that the tip is illuminated both directly and by a beam reflected from the sample surface and that the scattered light propagates towards the detector both directly and after being reflected from the surface [13] (i.e., $r$ is the THz far-field *p*-polarization reflectance of the sample at 60° incidence in our setup, including interferences in the photoexcited layer). Note that all the quantities, $s^{(m)}$, $\alpha^{(m)}$, and $r$, are generally frequency dependent (for the sake of the graphical clarity we omit marking this dependence in the formulae).

Altogether, Eq. (1) is a rather complex expression controlled not only by the response of the sample, but also by the tip geometry and settings of the SNOM (namely by the tip radius, minimum tip – sample distance, and the tapping amplitude) and by semi-empirical parameters (effective tip length $L$ and the $g$ factor which is usually set to $0.7e^{0.06i}$ according to the literature [13]).

Similarly to conventional spectroscopies, the detected signal depends on the intensity and shape of the incident pulse and on the instrumental response function. These are too difficult to evaluate, and it is thus necessary to establish a suitable reference. The standard possibility is normalization by the signal scattered from the unexcited sample; the analyzed scattered signal then reads

$$S^{(m)} \equiv \frac{s_{\text{exc}}^{(m)}}{s_{\text{gnd}}^{(m)}} = \frac{\alpha_{\text{exc}}^{(m)}}{\alpha_{\text{gnd}}^{(m)}} \frac{(1+r_{\text{exc}})^2}{(1+r_{\text{gnd}})^2} \equiv A^{(m)} \frac{(1+r_{\text{exc}})^2}{(1+r_{\text{gnd}})^2} \tag{2}$$

Here the subscript *exc* refers to the photoexcited sample and *gnd* to the unexcited sample (ground state), and $A^{(m)}$ expresses the ratio of tip polarizabilities in the excited and ground state. With the scattering-type SNOM, it is possible to devise also a self-referenced scheme, in which one analyzes the ratio of scattered signals belonging to different tip tapping harmonics, $m$ and $m'$, which is in fact equivalent to the ratio of the polarizabilities:

$$\frac{s_{\text{exc}}^{(m')}}{s_{\text{exc}}^{(m)}} = \frac{\alpha_{\text{exc}}^{(m')}}{\alpha_{\text{exc}}^{(m)}} \tag{3}$$

The potential advantage of this approach is the elimination of the sample reflectance; such strategy could be particularly interesting for samples with complex surfaces where it may be difficult if not impossible to estimate the reflectance. On the other hand, any systematic error in the signal depending on the harmonics order would translate into a systematic error in the ratio (3). Such an issue should be eliminated upon normalization by appropriate ground-state signals:

$$X^{(mm')} = \frac{\frac{s_{exc}^{(m')}}{s_{gnd}^{(m')}}}{\frac{s_{exc}^{(m)}}{s_{gnd}^{(m)}}} = \frac{A^{(m')}}{A^{(m)}} \qquad (4)$$

This expression is independent of the sample reflectance, too.

## 3.2. Sensitivity study

We examine the sensitivity of these three approaches for the model case closely related to the experiments with GaAs and InP samples presented later in this paper. We consider that the sample consists of a homogeneous photoexcited layer with thickness $d$ and properties of a high-resistivity GaAs excited at 780 nm (see the parameters in the caption of Fig. 1) on top of a semi-infinite unexcited substrate. The dielectric permittivity of the photoexcited sample (needed for the calculation of the SNOM response [13, 14]) is modelled by the Drude formula:

$$\varepsilon_{exc} = \varepsilon_{gnd} \underbrace{- \frac{1}{2\pi i f \varepsilon_0} \cdot \frac{n_{exc} e_0^2}{m_{eff}} \cdot \frac{\tau}{1 - 2\pi i f \tau}}_{\Delta \varepsilon} \qquad (5)$$

where $\varepsilon_{gnd}$ is the ground state permittivity, $\varepsilon_0$ is the vacuum permittivity, $\Delta \varepsilon$ is the change of permittivity due to photoexcitation, $f$ is the frequency of the THz light, $n_{exc}$ is the density of photoexcited carriers, $\tau$ is their scattering time and $m_{eff}$ is their effective mass. For semiconductors, it is more natural to describe their photoinduced response in terms of their photoconductivity:

$$\Delta \sigma = -2\pi i f \varepsilon_0 \Delta \varepsilon \qquad (6)$$

or sheet photoconductivity:

$$\Delta \Sigma = d \Delta \sigma. \qquad (7)$$

Since we encounter scattering times $\tau$ of about 150 fs and our central frequency $f$ is about 1 THz, the term $2\pi f \tau$ in Eq. (5) is close to 1. In turn, our investigations are concentrated on the regime where the magnitudes of the real and imaginary part of the photoconductivity are comparable to each other (the same holds for the photoinduced change in the permittivity).

In Fig. 1, we plot the scattered signal normalized by that from the unexcited sample [$S^{(2)}$, Eq. (2)] along with its decomposition into the individual terms (ratio of the tip polarizabilities $A^{(2)}$ and ratio of far-field reflectances). We investigate the dependence on the carrier density and photoexcited layer thickness, which are two typical important parameters to be extracted from the measurements. While the most important dependence enters through the ratio of the tip polarizabilities $A^{(2)}$ [Fig. 1(b)], the far-field reflectance term $(1 + r_{\text{exc}})^2/(1 + r_{\text{gnd}})^2$ starts to play a considerable role for parameters corresponding to the upper right corner of Fig. 1(c), i.e., for the highest sheet conductivity values in the range presented in Fig. 1.

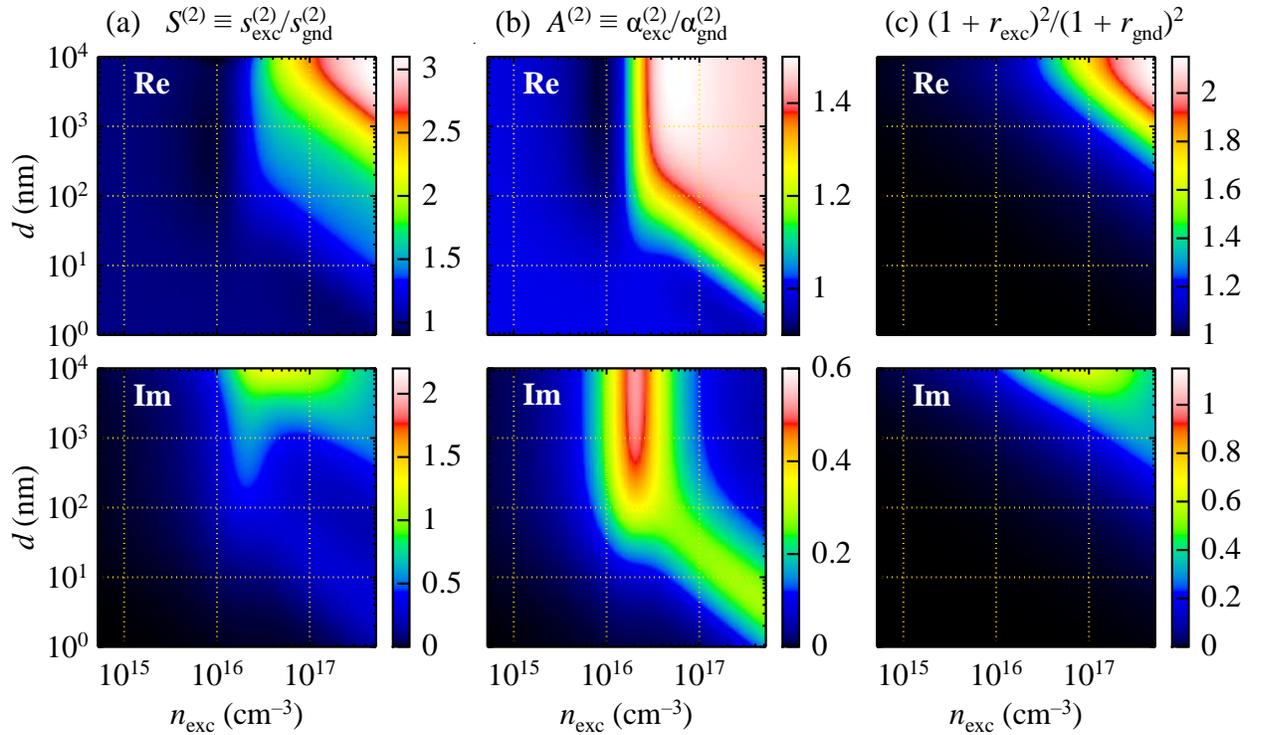

Figure 1: (a) The ratio of the complex scattered signals in the excited and ground state demodulated at the 2nd harmonics of the tip tapping frequency and its decomposition into (b) ratio of polarizabilities (near-field contribution) and (c) far-field reflectance contribution. Model parameters: $f$ = 1 THz, tip tapping amplitude $\Delta H$ = 150 nm, tip radius $R$ = 40 nm, effective tip length $L$ = 300 nm and $g = 0.7e^{0.06i}$. Properties of GaAs: $\varepsilon_{\text{gnd}} = 13$, electron scattering time $\tau$ = 150 fs and electron effective mass $m_{\text{eff}} = 0.067 m_{\text{e}}$.

We represent the sensitivity of the measurement by the term $|S^{(m)} - 1|$ which corresponds to the relative change due to the photoexcitation; an analogic expression, $|A^{(m)} - 1|$, is used to characterize the measurement sensitivity with respect to the near field polarizability. The calculated data for the signal demodulated at $2\Omega_{\text{tip}}$ in our model GaAs sample is shown in Fig. 2(a) and (b). The contour lines show various levels of sensitivity in per cents. Roughly, their shape between 3 and 30% can be approximated by an essentially vertical half line for thicker layers and a diagonal half line with a slope of −1 in the log-log scale for thinner layers. This approximation is indicated in Fig. 2(b) by the red dashed line for the sensitivity of 10%. The same trend is observed also for the normalized tip polarizability factor shown in Fig. 1(b) where both the vertical and diagonal parts are clearly apparent. From these simple comparisons of Figs. 1 and 2 it follows that up to about 30%, the sensitivity is controlled by the near-field polarizability which provides here the main contribution to the signal. Fig. 2(c) and (d) show the real part of conductivity and sheet conductivity within the explored space of parameters; the borders of the 10% sensitivity are plotted again as dashed lines. It follows that, in order to observe a 10% change in the signal, a minimum conductivity change of Re $\Delta\sigma \gtrsim 0.5$ S/cm and a minimum sheet conductivity change of Re $\Delta\Sigma \gtrsim 4$ μS are required. The full diagonal line in Fig. 2(d) approximately indicates the onset of the far-field reflectance contribution to the signal (Re $\Delta\Sigma \gtrsim 1$ mS in our model case).

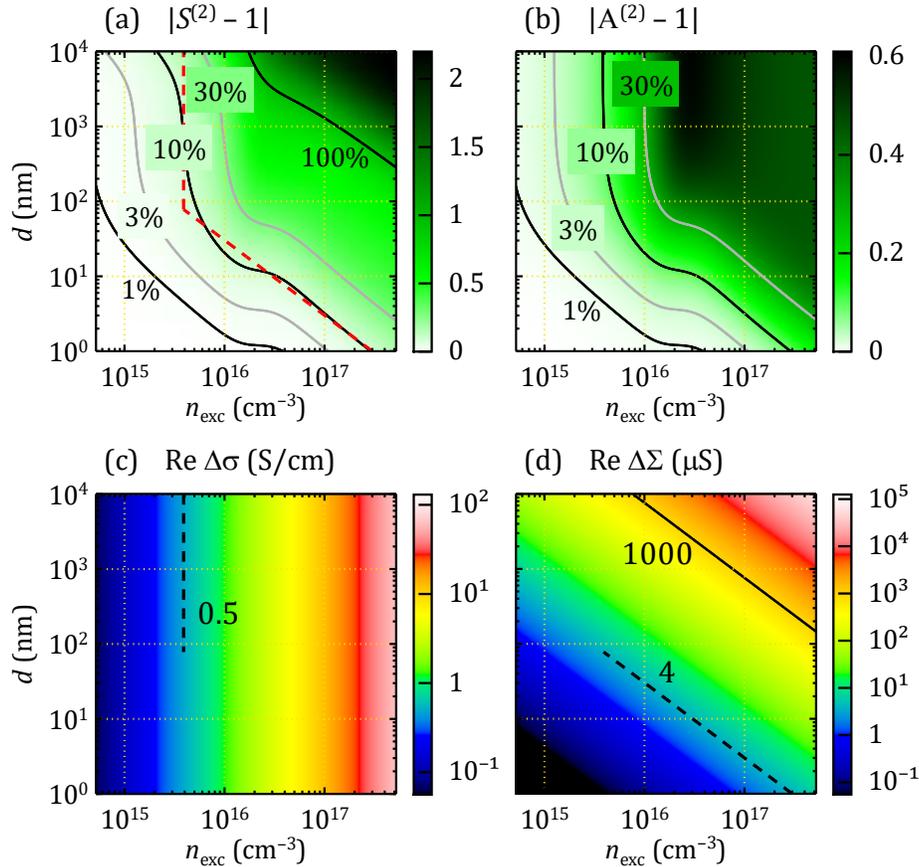

Figure 2: (a,b) Contour/colormap plots of the sensitivity of the signals (a) $|S^{(2)} - 1|$ and (b) $|A^{(2)} - 1|$ as a function of photocarrier concentration and film thickness. The red dashed line in (a) represents the approximation of the 10% sensitivity contour described in the text. (c,d) Plots relating the sensitivity to the corresponding real (c) photoconductivity and (d) sheet

photoconductivity of the photoexcited layer. The dashed lines correspond to the dashed segments in the plot (a) and express the minimum values of Re $\Delta\sigma \approx 0.5$ S/cm and Re $\Delta\Sigma \approx 4$ µS for achieving 10% sensitivity. The solid line (Re $\Delta\Sigma \approx 1$ mS) indicates the onset of the far-field reflectance contribution. Frequency of 1 THz is considered in all panels; model parameters are the same as in Fig. 1.

## 3.3. Correlation of the retrieved parameters of photoexcited layer

Besides the sensitivity issues, the retrieval is complicated by the fact that the inverse mapping of the complex signal $S^{(m)}$ to $n_{\text{exc}}$ and $d$ might be ambiguous in some regions of the parameter space $(n_{\text{exc}}, d)$. To learn more about the inverse mapping, it is convenient to investigate the sensitivity of the retrieved values $(\ln n_{\text{exc}}, \ln d)$ on the complex signal $S^{(m)}$. In our model case, we assume that the absolute errors in the real ($\delta S_{\text{re}}$) and imaginary part ($\delta S_{\text{im}}$) of the complex signal $S^{(m)}$ (or any other signal later in the manuscript) reach the same amplitude and that they are not correlated; the errors in $S^{(m)}$ are then represented by a circle with the radius given by the error amplitude $|\delta S|$ (Fig. 3a). For reasonably small $|\delta S|$, linear expansion applies and the errors $\delta(\ln n_{\text{exc}})$ and $\delta(\ln d)$ (which are equivalent to the relative errors $\delta n_{\text{exc}}/n_{\text{exc}}$ and $\delta d/d$) can be expressed using the matrix transformation:

$$\begin{pmatrix} \delta \ln n_{\text{exc}} \\ \delta \ln d \end{pmatrix} = \begin{pmatrix} n_{\text{exc}} \dfrac{\partial \operatorname{Re} S}{\partial n_{\text{exc}}} & d \dfrac{\partial \operatorname{Re} S}{\partial d} \\ n_{\text{exc}} \dfrac{\partial \operatorname{Im} S}{\partial n_{\text{exc}}} & d \dfrac{\partial \operatorname{Im} S}{\partial d} \end{pmatrix}^{-1} \begin{pmatrix} \delta S_{\text{re}} \\ \delta S_{\text{im}} \end{pmatrix} \qquad (8)$$

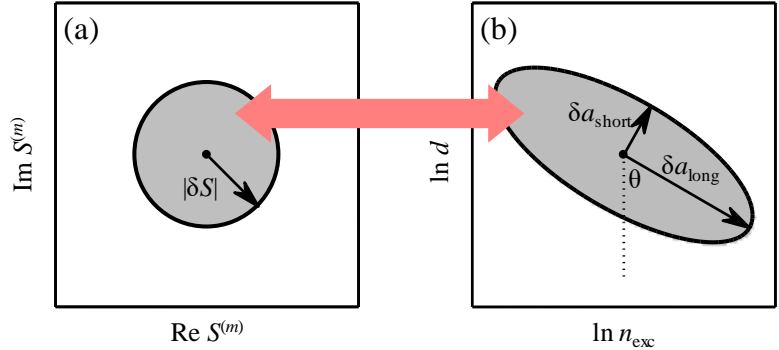

Figure 3: Illustration of the inverse mapping of the measured signal $(\operatorname{Re} S, \operatorname{Im} S)$ to the retrieved parameters $(\ln n_{\text{exc}}, \ln d)$.

In general, the initial circle transforms into an inclined ellipse (Fig. 3b), called characteristic error ellipse in the following. The lengths of its semi-axes $\delta a_{\text{short}}$ and $\delta a_{\text{long}}$ are given by the following linear combinations:

$$\begin{aligned} \delta a_{\text{short}} &= \cos\theta\, \delta\ln n_{exc} + \sin\theta\, \delta\ln d \\ \delta a_{\text{long}} &= -\sin\theta\, \delta\ln n_{exc} + \cos\theta\, \delta\ln d, \end{aligned} \qquad (9)$$

and they represent the errors in the determination of these linear combinations. It implies that the errors in the quantities $\ln n_{exc}$ and $\ln d$ are in general correlated and that the linear combination along the shorter semi-axis can be determined with a better accuracy than the linear combination along the longer semi-axis.

We applied this approach to our model photoexcited semiconductor layer and the results represented by $\delta a_{short}$ and $\delta a_{long}$ are shown in Fig. 4, together with the inclination of the ellipse given by the angle $\theta$. Next, we discuss several important regions in the ($\ln n_{exc}$, $\ln d$) space denoted by numbers ①–④ in Fig. 4, where $\delta a_{short}$ is quite small (dark blue regions in Fig. 4a), i.e., where the resulting error might be reasonably small for one of the parameters $\ln n_{exc}$, $\ln d$, or for their appropriate linear combination.

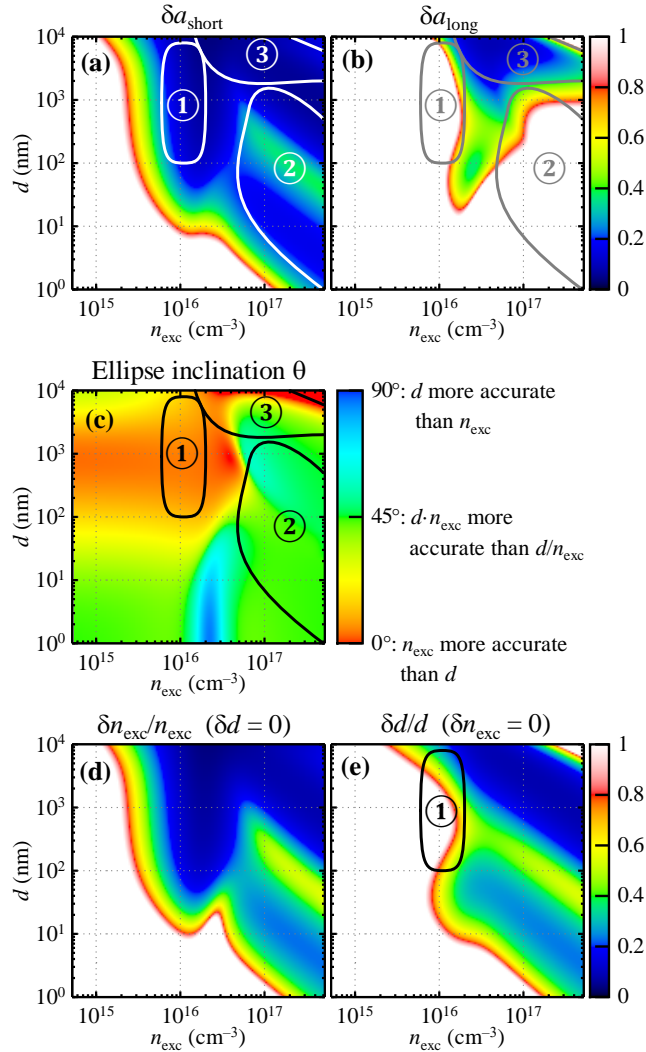

Fig. 4. (a), (b) Colormap plots of short (a) and long (b) axes of the error ellipse in the ($\ln n_{exc}$, $\ln d$) space assuming the absolute error in the SNOM signal $S^{(2)}$ of $|\delta S^{(2)}| = 0.05$, which seems a reasonable estimate for a typical experiment. (c) Inclination of the error ellipse as defined in Fig. 3, and, in turn, also the correlation between the relative errors in $n_{\text{exc}}$ and $d$ ($\delta n_{\text{exc}}/n_{\text{exc}}$ and $\delta d/d$). (d,e) Relative errors in $n_{\text{exc}}$ and $d$ calculated under an additional assumption that (d) $d$ is *a priori* known ($\delta d = 0$) or (e) $n_{\text{exc}}$ is *a priori* known ($\delta n_{\text{exc}} = 0$). In white regions the calculated error is 100% or larger. All the plots were calculated for the frequency of 1 THz. The numbered regions are described in the text.

In region ① the characteristic error ellipse is nearly vertical ($\theta \approx 0°$) and reasonably narrow. Consequently, in this region, the carrier density can be determined accurately (5% error in the signal amplitude translates into less than 20% error in $n_{\text{exc}}$). However, $\delta a_{\text{long}}$ acquires very high values (since $|d \cdot \partial S^{(2)}/\partial d| \ll |n_{\text{exc}} \cdot \partial S^{(2)}/\partial n_{\text{exc}}|$); therefore, it is practically impossible to determine the photoexcited layer thickness.

In region ② the characteristic error ellipse is inclined by 45° counterclockwise indicating that the sum $\ln n_{\text{exc}} + \ln d$ (or, equivalently, the product $n_{\text{exc}} \cdot d$) can be retrieved reliably, while the difference $\ln n_{\text{exc}} - \ln d$ (or, equivalently, the ratio $n_{\text{exc}}/d$) cannot. Thus, this region can yield the sheet photoconductivity of the photoexcited layer, similarly as the conventional far-field transient THz transmission experiments on thin films.

Region ③ is characterized by the best sensitivity for an independent retrieval of $n_{\text{exc}}$ and $d$. Note that this is the region where the reflectance term contributes significantly to the signal as observed in Fig. 1(c) and it thus seems that it is the interference in the thin photoexcited layer which is needed to facilitate the deconvolution of the carrier density and photoexcited layer thickness.

In comparison, we show in Fig. 4(d) the relative error in $n_{\text{exc}}$ assuming that $d$ is exactly known (e.g., the photoexcited layer thickness is determined independently from the absorption coefficient of the material at the excitation wavelength) and, similarly, in Fig. 4(e), we show the relative error in $d$ assuming that $n_{\text{exc}}$ is exactly known. The accessible regions of $n_{\text{exc}}$ and $d$ become significantly larger under these conditions and they both resemble the area of small values of $\delta a_{\text{short}}$ depicted in Fig. 4(a) with the exception of region ① where the characteristic ellipse is nearly vertical and therefore $d$ cannot be determined.

In brief, Figs. 4(a) and (c) together show that it is possible to retrieve either the conductivity or the sheet conductivity of the sample for a rather large space of parameters. Fig. 4b then indicates the region where *both* $n_{\text{exc}}$ and $d$ can be retrieved with an acceptable accuracy.

## 3.4. Frequency dependence

So far, we analyzed the signal behavior at the frequency of 1 THz. These conclusions can be easily generalized to an arbitrary frequency when the conductivity is dispersion free and the layer thickness is small. In such a case, it is the $1/f$ dependence in Eq. (5) which controls the scattered signal. Scaling of the frequency is then equivalent to a reciprocal scaling of the conductivity or carrier density $n_{\text{exc}}$; i.e., all patterns drawn in the ($\ln n_{\text{exc}}$, $\ln d$) plane just shift in the $-\ln n_{\text{exc}}$ direction (as seen for $d \lesssim 100$ nm in Fig. 5). More importantly, it becomes obvious that the analysis of a broadband spectrum enhances the possibility of reaching a region with a good sensitivity and thus allow determination of the parameters by means of a frequency-dependent fit. These findings essentially underline the advantage in measuring the scattered signal in a broad spectral range.

In a general case, the frequency dependence is more complicated due to the dispersion of the conductivity. Namely, the imaginary part of the conductivity exceeds the real part at high frequencies, which leads to an enhancement of the imaginary part of the signal $S^{(2)}$ as observed in Fig. 5 for 2 THz. Furthermore, for optically thicker photoexcited layers interferences of the THz beam in the layer become important and further distort the pattern.

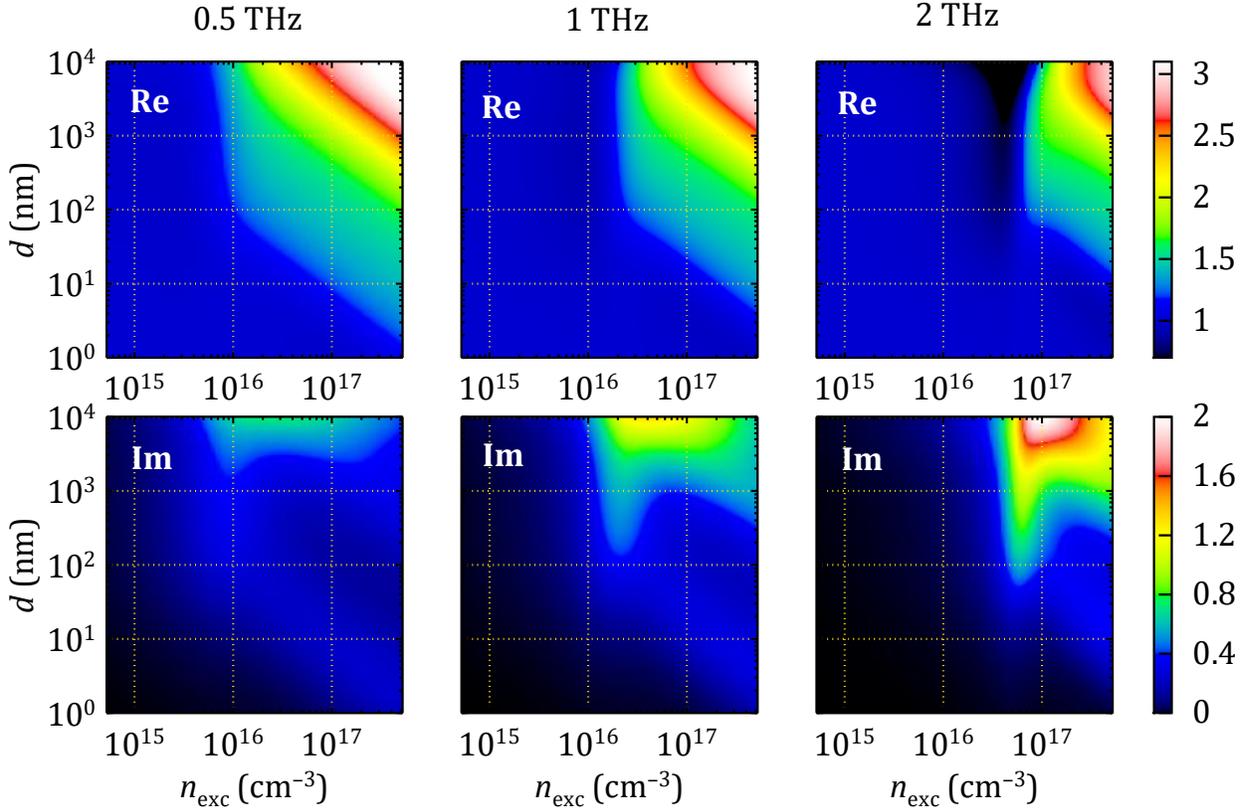

Figure 5. Frequency dependence of the signal $S^{(2)} \equiv s^{(2)}_{\text{exc}}/s^{(2)}_{\text{gnd}}$. Model parameters are the same as in Fig. 1.

## 3.5. Notes on the signals at higher harmonics

Comparison of the ratios $S^{(m)}$ for the signal demodulation at different harmonics $m = 2,3,4$ shows qualitatively the same behavior for each $m$ (Fig. 6). Quantitative changes are significant and can be visualized using contour lines. We observe namely a considerable vertical shift of the contour line in the high-signal region, i.e., a shift towards thinner photoexcited layer upon increasing the harmonic order. In other words, by using higher harmonics we reach a better sensitivity for thinner layers. Such a finding was recently pointed out in an experimental study of ultrafast carrier dynamics in GaAs nanobars [15]. Our current theoretical analysis thus further confirms the fact that signal demodulated at higher harmonics comes from a thinner layer below the sample surface, and thus measurements using a set of harmonics in principle carry information about the depth profile of the photoconductivity.

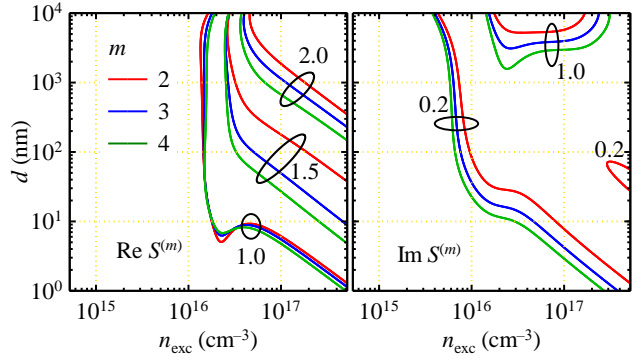

Fig. 6. Selected contour lines of the real and imaginary part of the ratio $S^{(m)}$ at 1 THz for the demodulation at 2nd, 3rd and 4th harmonics of the tip tapping frequency.

Unlike the signal $S^{(2)}$ examined so far, the signal $X^{(2,3)}$ defined by Eq. (4) cancels the reflectivity including the interferences of the incoming and scattered beams in the photoexcited layer, and it is thus somewhat easier to interpret, see Fig. 7(a,b) where areas ④ and ⑤ are indicated. The signal in the region ④ (photoexcited layer thickness exceeding ~50 nm) is controlled almost solely by the carrier density close to the surface whereas the signal in the region ⑤ ($d \lesssim 50$ nm) is controlled almost solely by the sheet conductivity. It is thus the latter region in which the SNOM is selectively sensitive to what happens close to the tip, within the depth comparable with the tip radius. The drawback of this approach is a very limited sensitivity even in the "most promising" areas ④ and ⑤ [Fig. 7(c,d)]: a 5% noise in the measured signal $X^{(2,3)}$ translates into at least 40% error in the retrieved carrier density (region ④) and sheet carrier density (region ⑤), which is ~4 times worse than for the signals $S^{(2)}$. Possibility to independently retrieve both the photoexcited layer thickness and carrier concentration is rather theoretical; there is a very small gap between the regions ④ and ⑤ where in the best case, both quantities can be obtained with approx. an order of magnitude accuracy. The reason is that the scattered intensities $s_{\text{exc}}^{(2)}$ and $s_{\text{exc}}^{(3)}$ depend on $n_{\text{exc}}$ and $d$ in rather similar ways: their ratio is thus almost constant and only the minor differences are available for the exploitation of information from the signal.

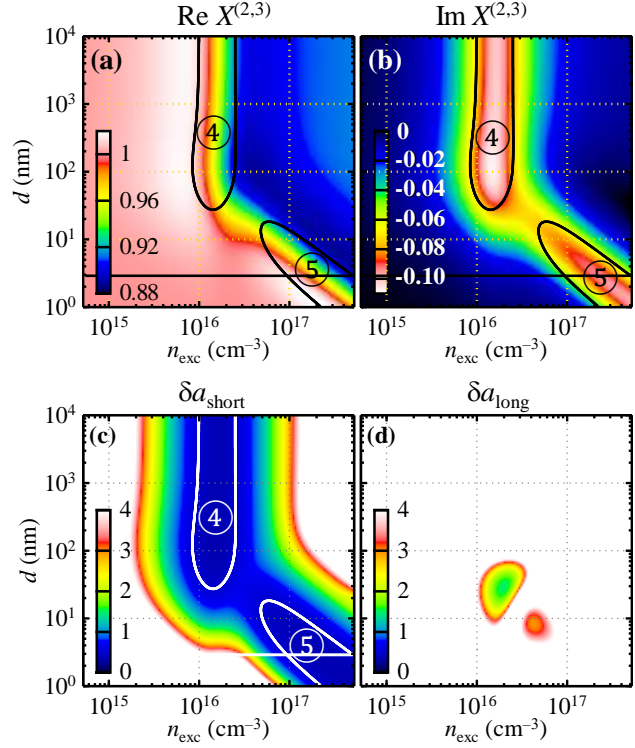

Fig. 7. (a,b) Real and imaginary part of the ratio of the complex scattered signal demodulated at the 2nd and 3rd harmonics, $X^{(2,3)}$. (c,d) Colormap plots of short and long axes of the error ellipse in the $(\ln n_{\text{exc}}, \ln d)$ space assuming the absolute error in the signal $X^{(2,3)}$ of 0.05. Note that the color scale indicates much larger values than in Fig. 4(a,b).

The character of the signal $s_{\text{exc}}^{(2)}/s_{\text{exc}}^{(3)}$ is almost identical to that of the signal $X^{(2,3)}$ (apart from the difference in the amplitudes) and the sensitivity to the errors is the same; for completeness, the plots are shown in the Appendix in Fig. A1. Due to the lack of any benefits, this strategy will not be used for the analysis of experimental data.

## 4. Experimental results for GaAs

### 4.1. Signal $S^{(2)}$

The $S^{(2)}$ spectra measured on a bulk GaAs wafer (Fig. 8) exhibit a broad tip-plasmon resonance which red-shifts as the density of photoexcited carriers decays with time after photoexcitation. Using far-field transient THz transmission spectroscopy, we verified that the lifetime of photoexcited charges in GaAs is ≳100 ps, which implies that all photocarriers recombine before the next pump laser pulse impinges on the sample. Immediately after photoexcitation, the spatial carrier profile is thus a simple exponential with $1/e$ decay length of 750 nm in GaAs. Since the electron mobility in GaAs (~7000 cm$^2$V$^{-1}$s$^{-1}$) is much higher than the hole mobility (~400 cm$^2$V$^{-1}$s$^{-1}$), electrons rather than holes are probed.

For the analysis of the near-field interaction in GaAs, we approximate the photoexcited area by a layer with a (known) thickness of 750 nm with constant electron density, lying on an infinitely thick non-photoexcited bulk. The unknown material parameters are thus the free electron density $n_{\text{exc}}$ and the mean electron scattering time $\tau$. There are also two instrumental parameters: the parameter $g$ (set to $g = 0.7e^{0.06i}$ according to [13]), and the effective tip length $L$ (fitted, single value kept constant within each set of pump-probe delays).

The free electron density decays exponentially with time [Fig. 9(a)]; practically the same evolution as well as absolute carrier density is deduced from signals demodulated at the second and third harmonics of the tip tapping frequency. This is coherent with the assessed sensitivity [Fig. 4(d)]: the combination of a close-to-micron thick photoexcited layer together with the observed carrier densities of $1\times10^{16}$ to $5\times10^{16}$ cm$^{-3}$ falls just to the region where the carrier density can be determined with almost the best accuracy. The carrier density obtained from the measured data immediately after photoexcitation is $\sim 5.5 \times 10^{16}$ cm$^{-3}$.

The excitation density can be independently estimated from the excitation pulse energy (9 pJ) and from the excitation spot diameter visible on the camera in the SNOM (~10 µm, see Fig. A2 in the Appendix) while accounting for the reflection losses on the semiconductor surface (~60% for the employed $s$ polarization and for the angle of incidence of the excitation beam of 60°). Assuming the penetration depth of 700 nm [16], we obtain $2.8\times10^{17}$ cm$^{-3}$, a five times larger value. It should be stressed that this is an order of magnitude estimate. Firstly, the determination of the excitation spot size is inaccurate since it is very small and there is no way how to resolve the actual beam profile at the sample surface; the visible spot borders may correspond to a rather arbitrary drop in the excitation intensity. Moreover, the excitation spot is rather inhomogeneous and currently we cannot control the exact positioning of the tip with respect to the spot and thus also the actually sensed carrier density. Keeping in mind these issues, we consider the correspondence between the carrier density from the fit of the SNOM spectra and from the excitation fluence still acceptable.

The free electron density [Fig. 9(a)] decays almost two times faster than the scattered SNOM signal at the maximum of the THz pulse [Fig. 8(c)], and the long-lived component is somewhat more pronounced. This emphasizes a complex dependence of the scattered signal on the carrier density: although the raw evolution of the scattered signal is measurable more easily, an accurate assessment of the carrier dynamics generally requires a careful analysis of the spectra.

It is interesting that from the scattered SNOM signal it is possible to deduce also a more subtle effect: the evolution of the electron scattering time with carrier density [Fig. 9(b)]. These results are consistent with carrier-density-dependence of the electron mobility [17] (which is directly proportional to the electron scattering time $\tau_s$) and which drops almost by a factor of two for densities reaching $5\times10^{16}$ cm$^{-3}$. Again, values retrieved from signals demodulated at the second and third harmonics are well consistent, mostly within 10% error.

Somewhat puzzling may seem the observed effect that the effective tip length depends on the harmonics of the tip taping frequency: we obtained 0.87 ± 0.07 µm for the second harmonics, and 1.9 ± 0.3 µm for the third harmonics. Both values are considerably shorter than the real AFM tip length (80 µm). This indicates that one has to consider the effective tip length as an instrumental parameter with (so far) unclear relation to the real geometry of the setup. Nevertheless, the consistency of the results retrieved from signals demodulated at different tip tapping harmonics indicate that one can live with this approximation. We also realized that the effective tip length retrieved from repeated pump-probe experiments differs for different tips and setup adjustments, which suggests that it should be determined upon each change of the experimental setup.

Finally, note that the ambipolar diffusion length reaches 550 nm in 150 ps (for ambipolar diffusion coefficient $D_a$ = 20 cm$^2$/s, [18]) and it thus exceeds two thirds of the initial excitation depth for the longest investigated pump-probe delays. We verified that assuming the photoexcited layer even as thick as 1000 nm affects the fitted material parameters by less than 5%, and it is thus possible to neglect the diffusion on this level of accuracy. This agrees with the results of the sensitivity study summarized in Fig. 4(c) which showed that the scattered SNOM signal is almost insensitive to the photoexcited layer thicknesses for the given experimental parameters.

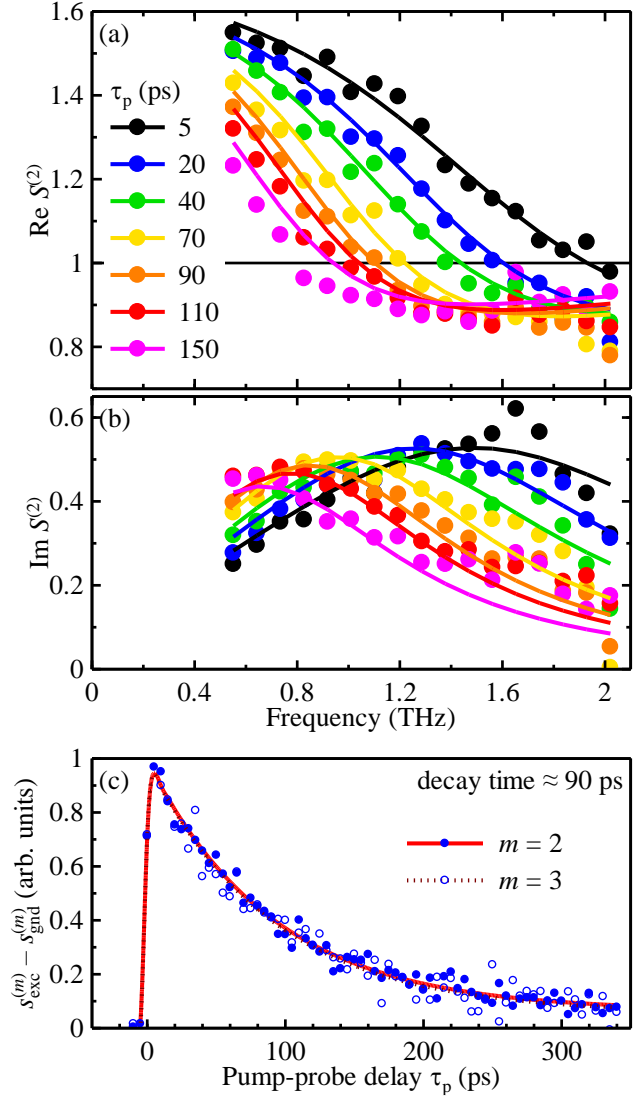

Figure 8: (a), (b) The measured real and imaginary part of the scattered SNOM signal normalized by that from the unexcited sample [$S^{(2)}$, Eq. (5)] at different times after photoexcitation $\tau_p$ in GaAs. Points represent the experimental data; lines are obtained from the finite-dipole model fit. (c) Time evolution of the amplitude of the photo-induced changes of the scattered signals from photoexcited GaAs demodulated at the 2nd and 3rd harmonics of the tip tapping frequency. Both signals are normalized to unity. Lines: fits by a mono-exponential (decay time ≈90 ps) with a long-lived component.

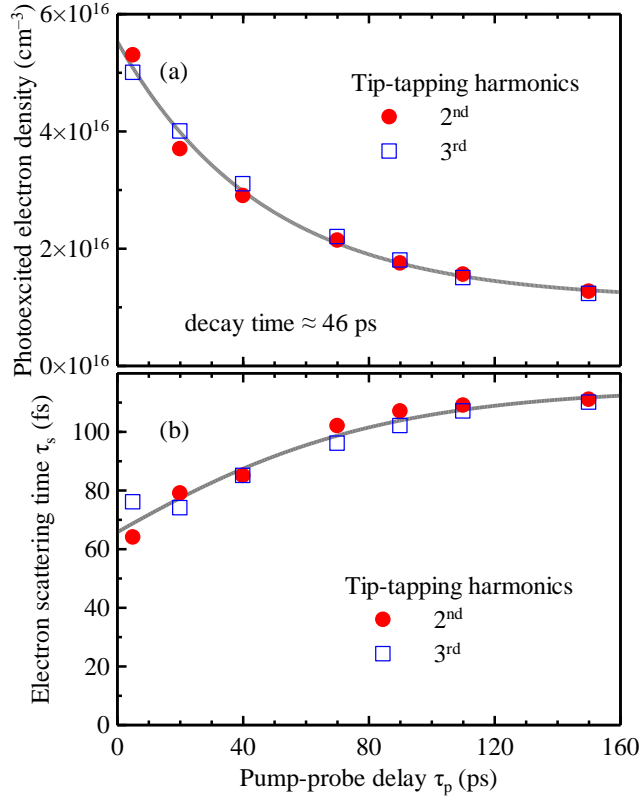

Figure 9: Evolution of the fitted (a) concentration of electrons and (b) electron scattering time in photoexcited GaAs. The line in (a) is a fit by a mono-exponential (decay time of ≈46 ps) with a long-lived component; the line in (b) serves to guide the eye only.

### 4.2. Signal $X^{(2,3)}$

Measurement of the spectra $X^{(2,3)}$ is more noisy, and namely the quality of the fit is much worse [Fig. 10(a,b)]. Despite that, both the value of the retrieved electron density and its evolution [Fig. 10(c)] match very well that found from the $S^{(2)}$ and $S^{(3)}$ spectra. Since the $X^{(2,3)}$ signal is insensitive to the layer thickness (Fig. 7), the photoexcited layer thickness was fixed to 750 nm in the fit; furthermore, the electron scattering time was fixed to 100 fs. For a good-quality fit, one needs to consider distinct effective tip lengths, 0.7 and 1.4 µm for the signals demodulated at the 2nd and 3rd harmonics of the tip tapping frequency, respectively. The difference from values obtained from $S^{(2)}$ and $S^{(3)}$ only underlines that this is an instrumental parameter so far.

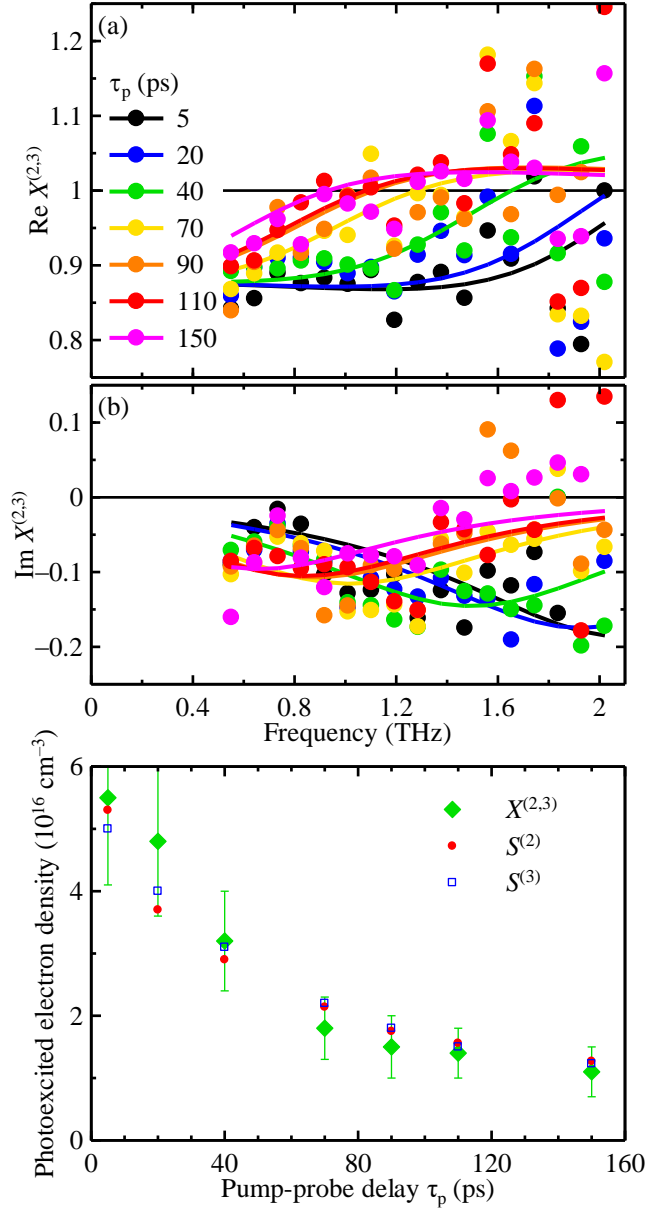

Figure 10: (a), (b) The measured real and imaginary part of the SNOM spectra $X^{(2,3)}$ at different times after photoexcitation $\tau_p$ in GaAs. Points represent the experimental data; lines are obtained from the finite-dipole model fit. (c) Evolution of the fitted concentration of electrons: comparison of results obtained from all three types of signals examined in the manuscript.

## 5. Experimental results for InP – signal $S^{(2)}$

The SNOM spectra measured on bulk InP (Fig. 11) are at first glance similar to those in GaAs: a broad plasmon resonance red-shifts as the density of photoexcited carriers decays with time after photoexcitation. The most prominent difference is the presence of the signal at a "negative" pump-probe delay, which in fact represents the state of the sample 10 ns after the arrival of the previous excitation pulse. When such long-lived electrons are present, one cannot neglect their redistribution in space due to the diffusion: The ambipolar diffusion coefficient of 10 cm$^2$s$^{-1}$ in InP [19] leads to the diffusion length of 3 µm during 10 ns in InP, which is an order of magnitude larger than the actual absorption depth of ~350 nm [16].

Furthermore, we observe that the decay of the $S^{(2)}$ signal [Fig. 11(c)] is significantly faster than the decay of the far-field transient transmittance (Fig. 13), which is proportional to the density of electrons integrated along the sample normal. This is strong evidence of a force pushing the electrons away from the surface, which is at the origin of the observed decrease of the scattered SNOM signal.

To model these two processes, we approximate the spatial distribution of free electrons using two layers (Fig. 12). The top layer with the thickness $d_{\text{exc}}$ contains a time-dependent free electron density $n_{\text{exc}}(\tau_{\text{p}})$: the layer thickness is set to the linear absorption depth in InP for pump-probe delays shorter than 1 ns [Fig. 12(a)], whereas it is allowed to expand to $d_{10\text{ ns}}$ for the longest measured pump-probe delay (10 ns) [Fig. 12(b)]. The bottom layer reaches down to the time-independent depth $d_{\text{long}}$ and we further consider that it contains a stationary free electron density $n_{\text{long}}$.

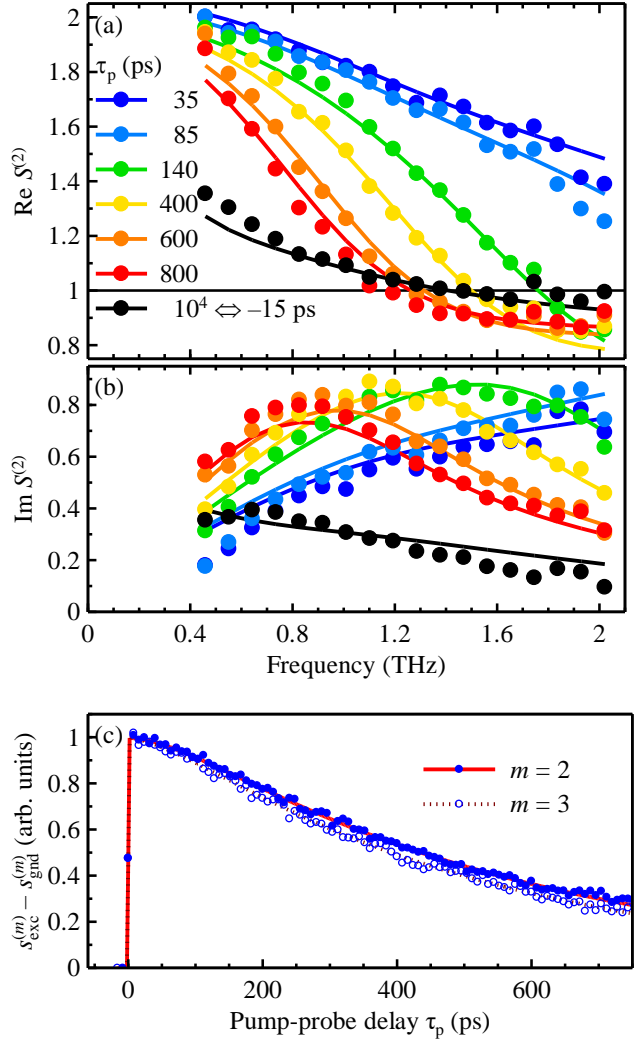

Figure 11: (a), (b) The measured real and imaginary part of the scattered SNOM signal normalized by that from the unexcited sample [Eq. (5)] at different times after photoexcitation $\tau_p$ in InP. Points represent the experimental data; lines are obtained from the finite-dipole model fit. (c) Time evolution of the amplitude of the photo-induced changes of the scattered signals from photoexcited InP demodulated at the 2nd and 3rd harmonics of the tip tapping frequency. The signals are normalized to unity.

The evolution of the electron density in the top layer (Fig. 13) obtained from the fit of the SNOM spectra shows faster dynamics than in GaAs, and considerably faster than the decay of the far-field signal representing the total electron density in the sample (including the charges in the deeper parts of the sample). This confirms the hypothesis that there is a driving force pushing electrons away from the surface: band-bending is a typical effect responsible for this transport [15, 20]. Dominance of surface recombination can be ruled out: it would decrease also the total density of electrons, which is not observed in the evolution of the far-field transient transmittance.

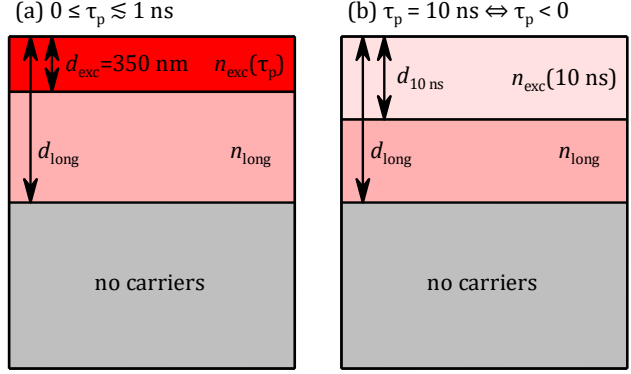

Figure 12: Scheme of the carrier distribution in the model for calculating the SNOM response in InP.

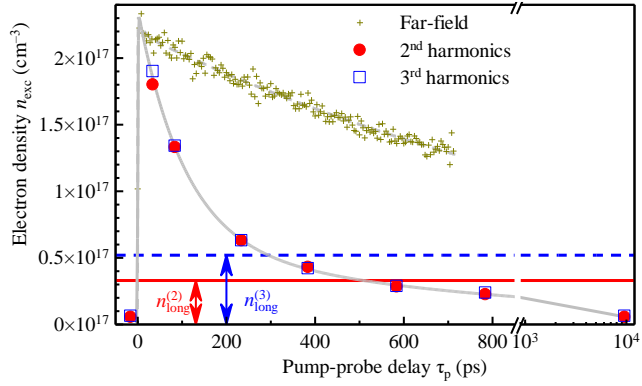

Figure 13: Evolution of the retrieved concentration of electrons in photoexcited InP. The red and blue lines indicate the density of the long-lived charges as obtained from signals demodulated and 2nd and 3rd harmonic. The grey lines serve to guide the eye only.

From the fits we found that the long-lived electrons spread down to $d_{long}$ = 6.0 µm away from the surface and that their density is $n_{long}$ = 3.3×10$^{16}$ cm$^{-3}$ (fit of the signal demodulated at the 2nd harmonics of the tip tapping frequency) or $n_{long}$ = 5.2×10$^{16}$ cm$^{-3}$ (fit of the 3rd harmonics). At 600 ps after photoexcitation, the electron density in the top layer decays below that of the long-lived electrons deeper in the sample, and it further drops down to almost zero at 10 ns; at this pump-probe delay, the top layer thickness expands to $d_{10\,ns}$ = 1.4 µm (we found the same value from the fits of signals demodulated at the 2nd and 3rd harmonics). These findings manifest band-bending close to the surface and therefore the presence of a force repelling electrons away from the surface [15]. Note also that the retrieved distances are close to the diameter of the excitation spot, which means that we are at the very limit where neglecting the lateral diffusion can be justified.

It is interesting to notice that here it was possible to independently retrieve both the carrier densities and the layer thicknesses with a reasonable accuracy. This benefit may be understood within the sensitivity analysis performed for the single layer system [Fig. 4(a,b)]: characteristic thicknesses around 6 μm and carrier densities on the level $3\times10^{16}$ cm$^{-3}$ fall right to the region where SNOM exhibits a rather good sensitivity to both quantities (mainly thanks to the surface reflectance). The density of the long-lived electrons $n_{\text{long}}$ is the only parameter for which the fitting of the 2$^{\text{nd}}$ and 3$^{\text{rd}}$ harmonics yields somewhat different values. We attribute this discrepancy to the fact that the spatial density profile, upon the combined action of the diffusion and the repelling force, is rather complex and cannot be reproduced within the approximation of the applied model involving two layers with homogeneous carrier density.

The fitted effective tip length (~0.37 μm for both harmonics) differs from the value found when dealing with GaAs and thus only underlines the conclusion that it is an uncertain instrumental parameter on the current level of the description. The fitted initial carrier density is about 5× lower than the estimate from the excitation fluence; this comparison is even better than for GaAs and can be explained in an identical way.

## 6. Summary

We analyzed theoretically the sensitivity of a terahertz scanning near-field optical microscope (THz-SNOM) for the determination of the free carrier density and thickness of the photoexcited layer with free carriers. The best measurable quantity for an independent determination of the carrier density and layer thickness is the ratio of the scattered signals from the excited and unexcited sample. The independent determination is possible namely for thicker layers thanks to the interferences of the incoming and scattered THz beam in the photoexcited layer. Analysis based on ratios of scattered signals at different tip-tapping harmonics is needed in the case of, e.g., structured sample surface when it is impossible to account for the surface reflectance affecting the incoming and scattered THz beam; however, in this case the results are considerably less accurate and allow the retrieval of either the carrier density or the layer thickness, but not both at the same time.

Using the developed strategies, we found out that GaAs is a prototypical semiconductor where the analysis of the SNOM spectra is rather straightforward. In contrast, InP is an example of a material which underlines the need of a very careful interpretation of the spectra measured by the SNOM. Band-bending is an important phenomenon which may influence the carrier distribution over several microns and at early time such as tens of picoseconds after photoexcitation. In our case, all these processes are directly seen in the spectra despite the fact that THz SNOM is typically associated with a sub-micron spatial resolution.

The possible presence of band-bending effects is an obvious complication for interpreting the THz-SNOM spectra of semiconductors, which is only emphasized by the dependence of the degree of the band-bending on the treatment of the surface. For example, we didn't see any manifestation of band bending in GaAs in this work, while an extremely rapid escape of electrons from the surface was observed was observed in MBE-grown GaAs layer which, moreover, underwent nanofabrication [15].

# Appendix – Supplementary figures

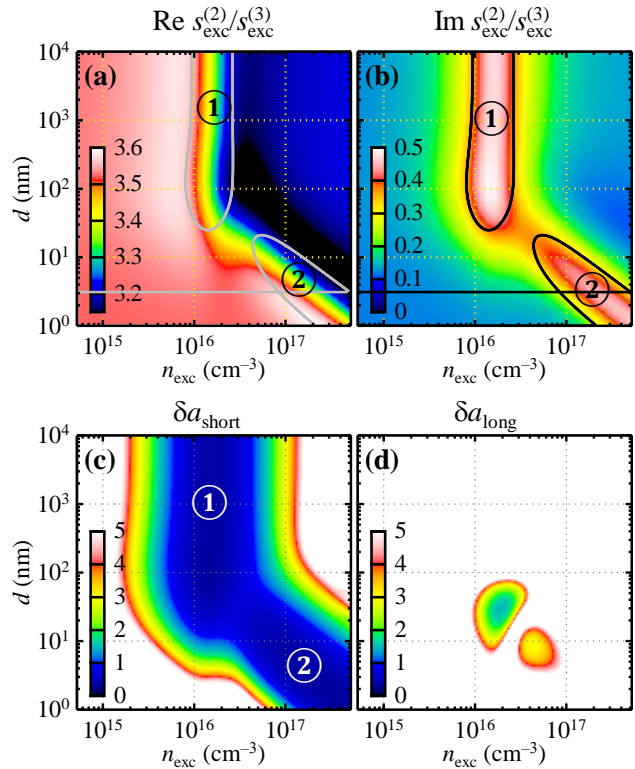

Figure A1. (a,b) Real and imaginary part of the ratio of complex scattered signal demodulated at the 2nd and 3rd harmonics of the tip tapping frequency, $s_{\text{exc}}^{(2)}/s_{\text{exc}}^{(3)}$. (c,d) Colormaps plots of short and long axes of the characteristic error ellipse in the $(\ln n_{\text{exc}}, \ln d)$ space assuming the relative error in the signal $s_{\text{exc}}^{(2)}/s_{\text{exc}}^{(3)}$ of 0.05. In the white regions the calculated error is 100% or larger. All the plots were calculated for the frequency of 1 THz and with the same model parameters as in Fig. 1.

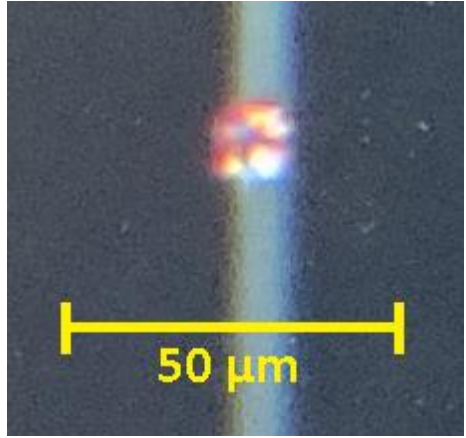

Figure A2. Photo of the excitation spot (red) hitting a groove in a glass. The profile of the excitation intensity is clearly neither homogenous, nor gaussian.

## Acknowledgment

This work was supported by the Czech Science Foundation (Project No. 23-05640S). This work was also co-financed by European Union and the Czech Ministry of Education, Youth and Sports (Project TERAFIT– CZ.02.01.01/00/22_008/0004594).

## References

[1] R. Ulbricht, E. Hendry, J. Shan, T. F. Heinz, and M. Bonn, Carrier dynamics in semiconductors studied with time-resolved terahertz spectroscopy, Rev. Mod. Phys. **83**, 543 (2011).

[2] J. Lloyd-Hughes and T.-I. Jeon, A review of the terahertz conductivity of bulk and nano-materials, Int. J. Infrared Millimeter Waves **33**, 871 (2012).

[3] P. Kužel and H. Němec, Terahertz spectroscopy of nanomaterials: a close look at charge-carrier transport, Adv. Opt. Mater. **8**, 1900623 (2020).

[4] C. J. Docherty and M. B. Johnston, Terahertz properties of graphene, J. Infrared. Milli. Terahz. Waves **33**, 797 (2012).

[5] C. A. Schmuttenmaer, Exploring dynamics in the far-infrared with terahertz spectroscopy, Chem. Rev. **104**, 1759 (2004).

[6] T. L. Cocker, V. Jelic, R. Hillenbrand, and F. A. Hegmann, Nanoscale terahertz scanning probe microscopy, Nature Photonics **15**, 558 (2021).

[7] V. Jelic, S. Adams, M. Hassan, K. Cleland-Host, S. E. Ammerman, and T. L. Cocker, Atomic-scale terahertz time-domain spectroscopy, Nature Photonics **18**, 898 (2024).

[8]  F. Zenhausern, M. P. O'Boyle, and H. K. Wickramasinghe, Apertureless near-field optical microscope, Appl. Phys. Lett. **65**, 1623 (1994).

[9]  F. Keilmann and R. Hillenbrand, Near-field microscopy by elastic light scattering from a tip, Phil. Trans. R. Soc. Lond. A **362**, 787 (2004).

[10] A. J. Huber, F. Keilmann, J. Wittborn, J. Aizpurua, and R. Hillenbrand, Terahertz near-field nanoscopy of mobile carriers in single semiconductor nanodevices, Nano Lett. **8**, 3766 (2008).

[11] M. Eisele, T. L. Cocker, M. A. Huber, M. Plankl, L. Viti, D. Ercolani, L. Sorba, M. S. Vitiello, and R. Huber, Ultrafast multi-terahertz nano-spectroscopy with sub-cycle temporal resolution, Nature Photonics **8**, 841 (2014).

[12] B. Knoll and F. Keilmann, Enhanced dielectric contrast in scattering-type scanning near-field optical microscopy, Opt. Commun. **182**, 321 (2000).

[13] A. Cvitkovic, N. Ocelic, and R. Hillenbrand; Analytical model for quantitative prediction of material contrasts in scattering-type near-field optical microscopy, Optica **15**, 8550 (2007).

[14] B. Hauer, A. P. Engelhardt, and T. Taubner, Quasi-analytical model for scattering infrared near-field microscopy on layered systems, Opt. Express **20**, 13173 (2012).

[15] V. Pushkarev, H. Němec, V. C. Paingad, J. Maňák, V. Jurka, V. Novák, T. Ostatnický, and P. Kužel, Charge transport in single-crystalline GaAs nanobars: Impact of band bending revealed by terahertz spectroscopy, Adv. Funct. Mater. **32**, 2107403 (2022).

[16] D. E. Aspnes and A. A. Studna, Dielectric functions and optical parameters of Si, Ge, GaP, GaAs, GaSb, InP, InAs, and InSb from 1.5 to 6.0 eV, Phys. Rev. B **27**, 985 (1983).

[17] D. L. Rode and S. Knight, Electron Transport in GaAs, Phys. Rev. B **3**, 2534 (1971).

[18] B. A. Ruzicka, L. K. Werake, H. Samassekou, and H. Zhao, Ambipolar diffusion of photoexcited carriers in bulk GaAs, Appl. Phys. Lett. 97, 262119 (2010).

[19] Y. Song, L. Cao, B. D. Peng, G. Z. Song, Z. Q. Yue, J. M. Ma, L. Sheng, B. K. Li, and H. X. Wang, Investigation of an InP-based image converter with optical excitation, Rev. Sci. Instrum. **88**, 033109 (2017).

[20] J. Lloyd-Hughes, S. K. E. Merchant, L. Sirbu, I. M. Tiginyanu, and M. B. Johnston, Terahertz photoconductivity of mobile electrons in nanoporous InP honeycombs, Phys. Rev. B **78**, 085320 (2008).